\documentclass[12pt]{article}
\usepackage[pctex32]{graphics}
\begin{document}
\vspace{2cm}
\begin{center}
~
\\
~
{\bf  \Large On the Supergravity Description of  Wilson Loop in Non-commutative Dipole Field Theory}
\vspace{1cm}

                      Wung-Hong Huang\\
                       Department of Physics\\
                       National Cheng Kung University\\
                       Tainan, Taiwan\\

\end{center}
\vspace{2cm}
The Wilson loop in the non-commutative dipole field theory is re-examined within the framework of dual gravity description.    In contrast to the previous investigations, we let the dual string be moving along the deformed $S^5$ and find the exact expression of the interquark potential.  The potential shows a Coulomb behavior at all distance and does not have a minimum distance between quarks, which  exhibits in the static configuration.   After comparing the potential of  the static and moving configurations we find that while the dual string is static at long distance it will transit to a moving configuration at short distance.  We also analyze an electric dipole system and find that it shows a similar transition property.   Finally, we mention the unsuitable approximation made in the previous paper [hep-th/0701069] and find that the interquark potential in a gauge theory with a non-constant non-commutativity has a  Coulomb behavior at all distance. 
\vspace{3cm}
\begin{flushleft}
*E-mail:  whhwung@mail.ncku.edu.tw\\
\end{flushleft}


\newpage
\section{Introduction}
The expectation value of Wilson loop could be investigated in the dual string description [1] in the spirit of  AdS/CFT correspondence [2]. The expectation value of the rectangular Wilson loop operator in the dual string description is evaluated by  the partition function of a string whose worldsheet is bounded by a loop and along a geodesic on the $AdS_5 \times S^5$ with endpoints on $AdS_5$.

  The computational technique has been used to the investigate the Wilson loop in various non-commutative gauge theories.  First,  the dual supergravity description of Wilson loop for the non-commutative gauge theory with a constant non-commutativity was studied by Maldacena and Russo [3].   They showed that the string cannot be located near the boundary and we need to take a non-static configuration in which the quark-antiquark acquire a velocity on the non-commutative space.  The result shows that the interquark potential exhibits the Coulomb type behavior as that in commutative space.  

  Another interesting non-commutative gauge theory is the non-commutative dipole field theory [4].  In [5] Alishahiha and Yavartanoo had investigated the associated Wilson loop in the dual gravity description and seen that the end points of string can be fixed at the boundary of AdS.  After the calculation they found that when the distance between quark and anti-quark is much bigger then their dipole size the energy will show a Coulomb type behavior with a small correction form the non-commutativity.  In a previous paper [6]  we re-examined the problem and found that it exists a minimum distance between the quarks.  

  In this paper, contrast to our previous investigations, we will let the dual string be moving along the deformed $S^5$.   We obtain a simple exact expression of the interquark potential which shows the Coulomb behavior at all distance.  Thus, the moving string configuration does not have a minimum distance as that exhibits in the static configuration.   We compare the potential between the static and moving configurations and find that while dual string is static at long distance it will transit to a moving configuration at short distance.   Especially, we analyze an electric dipole system and find that it shows similar transition property.  Finally, we mention the unsuitable approximation made in our previous paper [6] and see that the interquark potential in a gauge theory with a non-constant non-commutativity shows a Coulomb behavior at all distance.   We also discuss some interesting physical properties of dual string configuration and Wilson loop in the these non-commutativity  theories.


Let us first consider the non-commutative dipole theory.   The string theory realization of the 4D non-commutative dipole field theory was found in [4].  The corresponding geometry is described by
$$ds_{10}^2 = U^2\left(- dt^2+ dx_1^2+ dx_2^2+{ dz^2\over 1+B^2U^2\sin^2\theta_1\sin^2\theta_2}\right)\hspace{4cm}$$
$$+ {1\over U^2} \left(dU^2+ U^2d\Omega_5^2-U^4B^2\sin^4\theta_1\sin^4\theta_2 {(a_3d\theta_3+a_4d\theta_4+a_5d\theta_5)^2\over 1+U^2B^2\sin^4\theta_1\sin^4\theta_2}\right). \eqno{(1)}$$
$$e^{2\Phi}= {1 \over  1+ U^4B^2\sin^4\theta_1\sin^4\theta_2},~~~
B_{z\theta_i}= - {a_i~U^2B^2\sin^4\theta_1\sin^4\theta_2 \over 1+U^2B^2\sin^4\theta_1\sin^4\theta_2 },\hspace{1.7cm}\eqno{(2)}$$
in which $a_3 \equiv \cos\theta_4 $, $a_4 \equiv - \sin\theta_3\cos\theta_3\sin\theta_4 $, and $a_5 \equiv \sin^2\theta_3\sin^2\theta_4$, where $\theta_i$ are the angular coordinates parameterizing the sphere $S^5$ transverse to the D3 brane.  Thus there is a nonzero B field with one leg along the brane worldvolume and others transverse to it.  The value $B$ in (2) is proportional to the dipole length $\ell$ defined in the ``non-commutative dipole product" : $\Phi_a (x) * \Phi_a (x) = \Phi_a (x-\ell_b/2) ~\Phi_b (x +\ell_a/2) $ for the dipole field $\Phi(x)$ [4]. 

  The supergravity description of the  Wilson loop on the  non-commutative dipole field theory was first investigated by Alishahiha and Yavartanoo [5].   They had found that the static string could be fixed at finite distance, in contrast to the non-commutative system studied by Maldacena and Russo [3].  Following their method we parameterize the static string configuration by
$$\tau=t,~~~~~~~U=\sigma,~~~~~~~z=z(\sigma),\eqno{(3)}$$
the Nambu-Goto action becomes
$$S= {1\over 2\pi}\int d\sigma d\tau \left(\sqrt{- det g} +B_{\mu\nu}\partial_\tau X^{\mu}\partial_\sigma X^{\nu}\right)={T\over 2\pi}\int d\sigma \sqrt{1+{U^4 (\partial_\sigma z)^2\over 1+ B^2U^2}},\eqno{(4)}$$
in which $T$ denotes the time interval we are considering and we have set $\alpha'=1$.  As the associated Lagrangian $({\cal L})$ does not explicitly depend on $z$ the function ${\partial{\cal L}\over \partial(\partial_\sigma z)}$ will be proportional to an integration constant, i.e.
$${\partial{\cal L}\over \partial(\partial_\sigma z)} ={{U^4 (\partial_\sigma z)\over 1+ B^2U^2} \over \sqrt{1+{U^4 (\partial_\sigma z)^2\over 1+ B^2U^2}}} = {U_0^2\over\sqrt{1+B^2U_0^2}},\eqno{(5)}$$
as at $U=U_0$ we have the property of  $(\partial_\sigma z) \rightarrow \infty$. From above relation we can find the function $(\partial_\sigma z)^2$
$$(\partial_\sigma z)^2 ={{1+B^2U^2\over U^4}\over {U^4\over U_0^4}{1+B^2U^2 \over 1+B^2U_0^2}-1} .\eqno{(6)}$$
Now, we put a quark at place $z=\sigma =-L/2$ and an anti-quark at $z=\sigma = L/2$,  thus
$$L = 2 \int_0^{L/2} d z = 2 \int_{U_0}^\infty dU (\partial_\sigma z)={2\over U_0}\int_1^\infty dy {\sqrt{1+y^2B^2U_0^2}\over y^2~\sqrt{y^4 {1+B^2U_0^2\over 1+y^2B^2U_0^2}-1}}$$
$$ = {2\over U_0}\int_0^1 dx {x^2 + B^2 U_0^2\over \sqrt{1 - x^2}\sqrt{1+B^2U_0^2 + x^2}}.\eqno{(7)}$$
Above relation implies that
$$L \approx  \{\begin{array} {ccc}
{2\over U_0}\int_0^1 dx {x^2 \over \sqrt{1- x^4}}=& {2\sqrt \pi \Gamma(3/4)\over U_0\Gamma(1/4)}& as~U_0\rightarrow~0,\\
2B \int_0^1 dx {1\over \sqrt{1-x^2)}}=& B \pi& as~U_0\rightarrow~\infty.
\end{array}\eqno{(8)}$$
Thus the interquark distant $L$ will asymptotically  approach to a constant $L_0\equiv B \pi$ as $U_0\rightarrow \infty$.   This indicates that it exists a minimum distance between the quark and anti-quark.

We can evaluate the interquark potential $H$ form the Nambu-Goto action (4) with a help of (6).  The formula is
$$H = {U_0\over \pi}\left[\int_1^\infty dy \sqrt{1+B^2U_0^2\over 1+y^2B^2U_0^2}\left({y^2\over \sqrt{y^4 {1+B^2U_0^2\over 1+y^2B^2U_0^2}}-1}-1\right)y^\epsilon-1\right]$$
$$ = {U_0 \sqrt{1+B^2U_0^2}\over \pi}\int_0^1 dx {x^{-\epsilon}\over x^2 \sqrt{1 - x^2}\sqrt{1+B^2U_0^2 + x^2}}.~~~~~~~~\eqno{(9)}$$
Here we have subtracted the infinity coming from the mass of W-boson which corresponding to the string stretching to $U=\infty$ [1]. 

  Now, using the relation (7) and (9) it is found that [5], when the distance between quark and antiquark is much bigger than their dipole size the interquark potential  is given by 
$$H(L) = -~ {1\over \pi~L}\left({\Gamma(3/4) \Gamma(1/2)\over \Gamma(1/4)}+ {1.5079 B^2\over L^2}\right),  ~~~~~as~~~L \gg L_0,\eqno{(10)}$$
The second term in the above equation shows that the dipole ($B$) has an effect to {\it attract} the quark and anti-quark to render their potential to be less than the commutative system. 

At short distance, i.e. $L \approx L_0$ we can from (9) find that at $U_0 \rightarrow \infty$  the interquark potential 
could be approximated as 
$$H = {U_0 \sqrt{1+B^2U_0^2}\over \pi BU_0}\int_0^1 dx {x^{-\epsilon}\over x^2 \sqrt{1 - x^2}\sqrt{1+ {1+x^2 \over B^2U_0^2}}}.~~~~~~~~$$
$$\approx {U_0 \over \pi}\int_0^1 dx {x^{-\epsilon}\left(1-{1+x^2\over 2B^2U_0^2}\right)\over x^2 \sqrt{1 - x^2}} = -{1\over 4 B^2 U_0}.\hspace{1.5cm}\eqno{(11)}$$
However, $L_0$ is not a really minimum distance, after a numerical analysis.  For a clear illustration we show in figure 1 the functions $L(U_0)$ and $H(U_0)$ at $B=1$ which is that by performing the numerical evaluation of (7) and (9).
\\
\\
\scalebox{1}{\hspace{1cm}\includegraphics{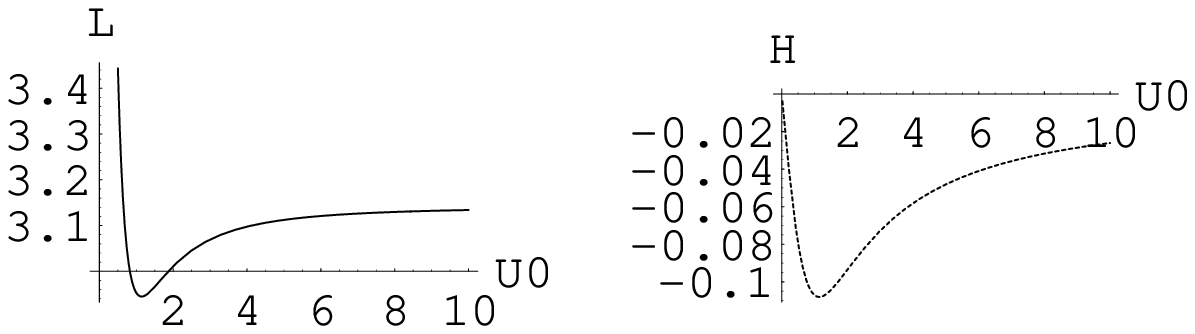}}
\\
\\
{\hspace{1cm} {\it Figure 1.  The function $L(U_0)$ and $H(U_0)$ at $B=1$.    We see that the minimum distance $L_0^*$ will be smaller than $L_0$.}
\\
\\
We thus see that the minimum distance $L_0^*$ will be smaller than $L_0$.

  In previous paper [6] we claimed that it appears a  repulsive force.   This is a wrong result of the unsuitable approximation.  It that paper we investigate the case of small value of $B$ under the limit of large $U_0$.  However, as it always appears a combination factor $B^2 U_0^2$ one does not know whether the value of $B^2 U_0^2$ is a small value or a large value.  This  unsuitable approximation was also adopted in our previous in considering the theory with a non-constant non-commutativity.  The corrected result will be presented later.

We now turn to investigate the dual string configuration which is moving with an angular velocity $\omega$ along the angular $\theta_3$.  Parameterizing the string configuration by
$$t = \tau,~~~~~~~U=\sigma,~~~~~~~z=z(\sigma),~~~~~~~\theta_3 = \omega \tau,\eqno{(12)}$$
the Nambu-Goto action becomes
$$S= {T\over 2\pi}\int d\sigma \sqrt{\left(1 - {\omega^2\over U^2 (1+B^2U^2)}\right)\left(1+{U^4 (\partial_\sigma z)^2\over 1+ B^2U^2}\right)} + {BU^2 \omega ~\partial_\sigma z\over 1+B^2 U^2},\eqno{(13)}$$
As the associated Lagrangian $({\cal L})$ does not explicitly depend on $z$ the function ${\partial{\cal L}\over \partial(\partial_\sigma z)}$ will be proportional to an integration constant, i.e.
$${\partial{\cal L}\over \partial(\partial_\sigma z)} ={\sqrt {1 - {\omega^2\over U^2 (1+B^2U^2)}}{U^4 (\partial_\sigma z)\over 1+ B^2U^2} \over \sqrt{1+{U^4 (\partial_\sigma z)^2\over 1+ B^2U^2}}}{BU^2 \omega \over 1+B^2 U^2} = {\omega \over B},\eqno{(14)}$$
as at $U \rightarrow \infty $ we have the property of  $(\partial_\sigma z) =0$ and $ U\cdot\partial_\sigma z =0$ to ensure that the end points of the string on the boundary has a finite distance.  From the above relation we can find the function $(\partial_\sigma z)^2$
$$(\partial_\sigma z)^2 ={\omega^2\over B^2}{1\over \left({1 - {\omega^2\over U^2 (1+B^2U^2)}}\right)U^8 - {\omega^2\over B^2} {U^4\over 1+B^2U^2}}.\eqno{(15)}$$
Using the property that $(\partial_\sigma z) \rightarrow \infty $ at $U =U_0 $ we find that
$$ \omega ^2 =  B^2~U_0^4 .\eqno{(16)}$$
Substituting this relation into (15) we have a simple relation
$$(\partial_\sigma z)^2 ={U_0^4\over U^4}{1\over U^4-U_0^4}.\eqno{(17)}$$
The interquark distance and potential could be calculated in the  similar way as that in the static case.   The results are
$$L = 2 \int_{U_0}^\infty dU (\partial_\sigma z)={2\over U_0}\int_1^\infty dy {1\over y^2~\sqrt{y^4 -1}} = {2\over U_0}\int_0^1 dx {x^2 \over \sqrt{1- x^4}} =  {1\over 2U_0} {\Gamma(3/4) \Gamma(1/2)\over \Gamma(1/4)}.\eqno{(18)}$$
$$H = {U_0\over \pi}\left[\int_1^\infty dy \left({y^2\over \sqrt{y^4 -1}}-1\right)y^\epsilon-1\right] = {U_0\over \pi }\int_0^1 dx {x^{-\epsilon} \over x^2 \sqrt{1- x^4}} =  -{U_0\over \pi} {\Gamma(3/4) \Gamma(1/2)\over \Gamma(1/4)}.\eqno{(19)}$$
The results are just those without a dipole field.  Note that the interquark potential calculated form above two equations is just (10) while without dipole term, thus the static string has less energy than a rotating string. 

 Comparing the potential between the static and moving configurations it shows that at large distance, i.e. $L \ge L_0^*$ the dual string is static.  However, at short distance, i.e. $L < L_0^*$  it could  appear as a moving configuration, which is a Coulomb potential and has an infinite attractive force at $L = 0$.  For a clear illustration we show in figure 2 the potential  $H(L)$.   The solid line is the potential of static string and dot line the moving string. 
\\
\\
\scalebox{1}{\hspace{5cm}\includegraphics{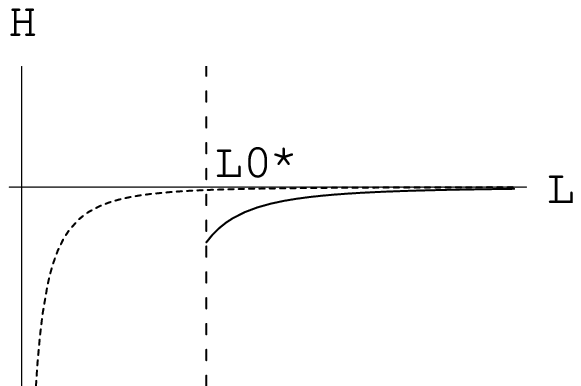}}
\\
\\
{\hspace{1cm} {\it Figure 2.  The interquark potential energy $H(L)$.  The solid line is the potential of static string and dot line the moving string.}
\\

Let us make following comments about our result.
\\
1.  According to the AdS/CFT duality the property of gauge theory  is calculated from the dual string on the AdS space.  The inter-quark force, which relates to  the gauge field property, is coded in the geometry of dual gravity background.  The existence of a minimum distance in the static dual string case reveals the {\it \bf geometry character} of the dipole deformed AdS background which lets the distance of endpoints $L$ be less then a non-zero value of $L_0^*$.
\\
2.  It is surprised that the moving string has the same result as that without dipole field.  The reason behind it may be argued  as following.  The dual string  in a background with  $B_{z \theta_3}$ field is somewhat analogous to the situation when a charged particle enters a region with a magnetic field.  Thus, the string will be rotating along $\theta_3$ with a constant angular momentum $\omega$ which is proportional to the strength of the NS-NS field, as shown in (16).   The configuration described in (17) has the {\bf binding energy} (it is negative) from B field which {\bf just be canceled by the kinetic energy} (it is positive) from the moving.  Thus the moving dual string does not depend on the  value of dipole field and we have the same result as that without dipole field.
\\
3.  As the appearance of a minimum distance in the static configuration seems quite strange we will analyze in below a electric dipole system and see that  it could show the similar property.  

 Consider the electric dipole with dipole moment $\vec p=  q\vec B$, in which $B$ is the length of the dipole.  Now, let the one dipole moment be located at $x=0$ and another dipole moment be located at $x$.   For the first configuration in which the two dipole moments are parallel to the $x-$axis the system has electric energy  
$$ H_{\parallel}(x) = {2q^2\over |x|} - {q^2\over |x-B|}-{q^2\over |x+B|} \approx \{\begin{array} {cc}
-\infty & as ~x\rightarrow~B\\
-{ 2q^2B^2\over |x^3|}& as~x \rightarrow~\infty.
\end{array}.\eqno{(20a)}$$
For the second configuration in which the first (second) dipole moment  is parallel (anti-parallel)  to the $y-$axis the system has electric energy  
$$ H_{\perp}(x) = {2q^2\over \sqrt{x^2+B^2}} - {2q^2\over |x|} \approx \{\begin{array} {cc}
-\infty & as ~x\rightarrow~0\\
-{q^2B^2\over |x^3|}& as~x \rightarrow~\infty.
\end{array}.\eqno{(20b)}$$
From above two equations we see that, at large distance,  the dipole will be at first configuration which at $x=x_0 \equiv B$ has an infinite attractive force, as described by the solid line in figure 3.   However, at short distance the dipoles will turn to the second configuration, as shown in the dot line plotted in figure 3.  
\\
\\
\scalebox{1}{\hspace{5cm}\includegraphics{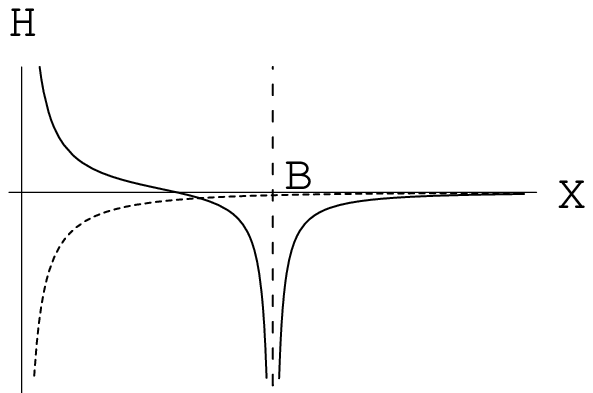}}
\\
\\
{\hspace{1cm} {\it Figure 3.  The inter-dipole potential energy $H(x)$.  The solid line is the potential of first configuration and dot line the second configuration .}
\\

Therefore one may conjecture that the static dual string and the rotating string will produce different intrinsic dynamics (something likes electric spin) arising from the dipole, which would then lead to the nontrivial behavior in the string action and thus the interquark potential.  Also, there is a phase  transition from static to moving configuration at short distance.  The details need furthermore analysis.
\\

Let us turn to the non-commutative theory with a non-constant non-commutativity.  The string theory realization of the 4D non-commutative field theory with a non-constant non-commutativity was found in [7].  The corresponding geometry is described by the Melvin-twist deformed $AdS_5\times S^5$ background  
$$ds_{10}^2 = U^2\left(- dt^2+ dr^2+{r^2 d\phi^2 + dz^2\over 1+ B^2r^2U^4}\right)+ {1\over U^2} \left(dU^2+ U^2d\Omega_5^2\right). \eqno{(21)}$$
$$e^{2\Phi}= {1 \over  1+ B^2r^2U^4},~~~~~C_{tr}= B^2rU^4, ~~~~C_{tr\phi z}= {rU^4 \over  1+ B^2r^2U^4}$$
$$B_{\phi z}= {Br^2U^4 \over  1+ B^2r^2U^4}.\hspace{7.5cm} \eqno{(22)}$$
\\
The ansatz for the dual string we will use is 
$$ t=\tau,~~~z=z(\sigma),~~~\phi = \omega \tau,~~~U=\sigma,\eqno{(23)}$$
and rest of the string position is constant in $\sigma$ and $\tau$. As that discussed by Maldacena and Russo [3],  the string cannot be located near the boundary and we need to take a non-static configuration in which the quark-antiquark acquire a velocity on the non-commutative space.  In the case with a constant non-commutativity the quark-antiquark will move along a direction with constant coordinate $x=x_0$.  However, in our case as the NS-NS field $B_{\phi z}$ described in (22) depends on a worldvolume coordinate $r$ the quark-antiquark will rotate along a constant coordinate  $r=r_0$ with a angular velocity $\omega$.  The physical interpretation is somewhat analogous to the situation when a charged particle enters a region with a radius-dependent magnetic field.  

 After choosing $r=r_0$ the Nambu-Goto action becomes
$$S={T\over 2\pi}\int d\sigma \left(\sqrt{\left(1- {r_0^2 \omega^2\over 1+ B^2r_0^2U^4}\right)\left(1+ {U^4 (\partial_\sigma z)^2 \over 1+ B^2r_0^2U^4}\right)}  ~+{\omega Br_0^2U^4~(\partial_\sigma z)\over 1+ B^2r_0^2U^4} \right),\eqno{(24)}$$
As the associated Lagrangian $({\cal L})$ does not explicitly depend on $z$ the function ${\partial{\cal L}\over \partial(\partial_\sigma z)}$ will be proportional to an integration constant, i.e.
$${\partial{\cal L}\over \partial(\partial_\sigma z)} ={\sqrt {1 - {r_0^2\omega^2\over (1+r_0^2 B^2U^4)}}{U^4 (\partial_\sigma z)\over 1+ r_0^2 B^2U^4} \over \sqrt{1+{U^4 (\partial_\sigma z)^2\over 1+ r_0^2 B^2U^4}}}+{r_0^2 BU^4 \omega \over 1+r_0^2 B^2 U^4} = {\omega \over B},\eqno{(25)}$$
as at $U \rightarrow \infty $ we have the property of  $(\partial_\sigma z) =0$.  From above relation we can find the function $(\partial_\sigma z)^2$
$$(\partial_\sigma z)^2 ={\omega^2\over B^2}{1\over \left({1 - {r_0^2\omega^2\over 1+B^2U^4)}}\right)U^8 - {\omega^2\over B^2} {U^4\over 1+r_0^2 B^2U^4}}.\eqno{(26)}$$
Using the property that $(\partial_\sigma z) \rightarrow \infty $ at $U =U_0 $ we find that
$$ \omega ^2 =  B^2~U_0^4 .\eqno{(27)}$$
Substituting this relation into (26) we have a simple relation
$$(\partial_\sigma z)^2 ={U_0^4\over U^4}{1\over U^4-U_0^4}.\eqno{(28)}$$
Above two equations are just (16) and (17).  The interquark distance and potential could be calculated in the  similar way and results are the same as eqs.(18) and (19) which are just those without a non-commutativity.\footnote {This property had been mentioned by Alishahiha et. al. [8] and we find that it is independent of  $r_0$.}

  Above results are exact and does not show a minimum distance,  in contrast to the previous paper [6].   The wrong result of our previous paper is due to the unsuitable approximation of small value of $B$ under the limit of large $U_0$, as it always appears a combination factor $B^2 U_0^2$ and one does not know whether the value of $B^2 U_0^2$ is a small value or a large vale. Note that the dual moving string has the same result as that without non-commutativity.  The reason behind it may be argued as that in the dipole case, i.e. the string rotating  under $B$ field has a negative binding energy which will just be canceled by the kinetic energy of the moving.  Thus the moving dual string does not depend on the  value of non-commutativity and we have the same result as that in commutative space.

In conclusion, we have shown that the dual string description of Wilson loop in the gauge theory with a  non-constant non-commutativity shows the Coulomb behavior at all distance, as that in the gauge theory with or without a  constant non-commutativity [1,3].  For the Wilson loop in the non-commutative dipole field theory, we show that at large distance the dipole could produce an attractive force to render the static configuration to have less energy than a moving one.  However, the static configuration could only be shown when the  interquark distance is larger than a minimum distance $L_0^*$.  A rotating dual string configuration would then correspond to the Wilson loop as the interquark distance is less than $L_0^*$.  We have compared the property of nono-commutative dipole system to an electric dipole system and conjecture that the static dual string and the rotating string will produce different intrinsic dynamics (something likes electric spin) arising from the dipole.  Thus the intrinsic dipole of Wilson loop may have some nontrivial dynamics which could lead to the interesting behavior in the string action and thus the interquark potential.  Also, there is a phase  transition from static to moving configuration at short distance.  The details need furthermore analysis.

\newpage
{\bf  \Large References}
\begin{enumerate}
\item J.~M. Maldacena,  ``{W}ilson loops in large {N} field theories,''  Phys.   Rev. Lett.  80 (1998) 4859-4862 [hep-th/9803002]; S.-J. Rey and J.-T. Yee,  ``Macroscopic strings as heavy quarks in large  {N} gauge theory and anti-de {S}itter  supergravity,''   Eur. Phys. J.   C22 (2001) 379--394 [hep-th/9803001].
\item J.~M. Maldacena, ``The large {N} limit of superconformal field theories  and supergravity,''  Adv. Theor. Math. Phys.  2  (1998) 231-252  [hep-th/9711200]; E.~Witten, ``Anti-de Sitter space and holography,'' Adv.\ Theor.\ Math.\ Phys.\   2 (1998) 253 [hep-th/9802150]; S.~S.~Gubser, I.~R.~Klebanov and A.~M.~Polyakov, ``Gauge theory correlators from non-critical string theory,'' Phys.\ Lett.\ B 428 (1998) 105 [hep-th/9802109].
\item J. M. Maldacena and J. G. Russo,`` Large N Limit of Non-Commutative Gauge Theories," JHEP 9909 (1999) 025 [hep-th/9908134].
\item A. Bergman and O. J. Ganor,``Dipoles, Twists and Noncommutative Gauge Theory," JHEP 0010 (2000) 018 [hep-th/0008030] ;  K. Dasgupta and M. M. Sheikh-Jabbari, ``Noncommutative Dipole Field Theories," JHEP 0202 (2002) 002 [hep-th/0112064]; A. Bergman, K. Dasgupta, O. J. Ganor, J. L. Karczmarek, and G. Rajesh,``Nonlocal Field Theories and their Gravity Duals," Phys.Rev. D65 (2002) 066005 [hep-th/0103090].
\item M. Alishahiha and H. Yavartanoo,``Supergravity Description of the Large N Noncommutative Dipole Field Theories," JHEP 0204 (2002) 031 [hep-th/0202131].
\item Wung-Hong Huang, ``Dual String Description of  Wilson Loop in Non-commutative Gauge Theory,''  Phys. Lett. B647 (2007) 519 [hep-th/0701069 ]. 
\item A. Hashimoto and K. Thomas, ``Dualities, Twists, and Gauge Theories with Non-Constant Non-Commutativity," JHEP 0501 (2005) 033 [hep-th/0410123]; A. Hashimoto and K. Thomas, ``Non-commutative gauge theory on D-branes in Melvin Universes," JHEP 0601 (2006) 083 [hep-th/0511197]. 
\item M. Alishahiha, B. Safarzadeh, and H. Yavartanoo``On Supergravity Solutions of Branes in Melvin Universes," JHEP 0601 (2006) 153 [hep-th/0512036].
\end{enumerate}
\newpage
\vspace{2cm}
\begin{center}
~
\\
~
{\bf  \Large Erratum to ``Dual string description of Wilson loop in non-commutative gauge theory" [ Phys. Lett. B647 (2007) 519 ]}
\vspace{1cm}

                      Wung-Hong Huang\\
                       Department of Physics\\
                       National Cheng Kung University\\
                       Tainan, Taiwan\\

\end{center}
~
Phys. Lett. B  652 (2007) 388-389
\begin{flushleft}
*E-mail:  whhwung@mail.ncku.edu.tw\\
\end{flushleft}
~
~
In our previous paper [1] we consider the case of small value of non-commutativity $B$ which, however, could not be applied in the limit of large $U_0$ as  it always appears a combination factor $B^2 U_0^2$.  The corrected result without approximation will be presented here.  

For the case of non-commutative dipole field theory the exact expressions of interquark distance and potential are
$$L =  {2\over U_0}\int_0^1 dx {x^2 + B^2 U_0^2\over \sqrt{1 - x^2}\sqrt{1+B^2U_0^2 + x^2}}.\eqno{(1)}$$
$$H = {U_0\over \pi}\left[\int_1^\infty dy \left( \sqrt{1+B^2U_0^2\over 1+y^2B^2U_0^2}{y^2\over \sqrt{y^4 {1+B^2U_0^2\over 1+y^2B^2U_0^2}}-1}-1\right)-1\right].\eqno{(2)}$$
After numerical evaluation we show in figure 1 the functions $L(U_0)$ and $H(U_0)$ at $B=1$.
\\
\\
\scalebox{1}{\hspace{1cm}\includegraphics{figurelh.eps}}
\\
\\
{\hspace{1cm} {\it Figure 1.  The function $L(U_0)$ and $H(U_0)$ at $B=1$.    We see that there is  a minimum distance $L_0^*$.}
\\

We now turn to investigate the dual string configuration which is moving with an angular velocity $\omega$ along the angular $\theta_3$.  Parameterizing the string configuration by
$$t = \tau,~~~~~~~U=\sigma,~~~~~~~z=z(\sigma),~~~~~~~\theta_3 = \omega \tau,\eqno{(3)}$$
the Nambu-Goto action becomes
$$S= {T\over 2\pi}\int d\sigma \sqrt{\left(1 - {\omega^2\over U^2 (1+B^2U^2)}\right)\left(1+{U^4 (\partial_\sigma z)^2\over 1+ B^2U^2}\right)} + {BU^2 \omega ~\partial_\sigma z\over 1+B^2 U^2}.\eqno{(4)}$$
In a similar way, we can find that $ \omega  =  B~U_0^2$.  Using this relation the interquark distance and potential are found to be just those without a dipole field. We show in figure 2 the potential  $H(L)$ in which the solid line is the potential of static string and dot line the moving string. 
\\
\\
\scalebox{1}{\hspace{5cm}\includegraphics{figurehl.eps}}
\\
\\
{\hspace{1cm} {\it Figure 2.  The interquark potential energy $H(L)$.  The solid line is the potential of static string and dot line the moving string.}
\\

 Comparing the potential between the static and moving configurations it shows that at large distance, i.e. $L \ge L_0^*$ the dual string is static.  However, at short distance, i.e. $L < L_0^*$  it could  becomes  a moving configuration. 

For the case of the non-commutative theory with a non-constant non-commutativity, the exact interquark distance and potential could be calculated in the  similar way.  (There is only the dual moving string solution.) The results are  just those without a non-commutativity, which had been mentioned by Alishahiha et. al. [2].

~

\begin{enumerate}
\item Wung-Hong Huang, ``Dual String Description of  Wilson Loop in Non-commutative Gauge Theory,''  Phys. Lett. B647 (2007) 519 [hep-th/0701069 ]. 
\item M. Alishahiha, B. Safarzadeh, and H. Yavartanoo``On Supergravity Solutions of Branes in Melvin Universes," JHEP 0601 (2006) 153 [hep-th/0512036].
\end{enumerate}
\end{document}